# High-performance Pockels-effect modulation and switching in silicon-based GaP/Si, AlP/Si, ZnS/Si, AlN/3C-SiC, GaAs/Ge, ZnSe/GaAs, and ZnSe/Ge superlattice-on-insulator integrated circuits

FRANCESCO DE LEONARDIS[1*], RICHARD SOREF[2]

[1]*Francesco De Leonardis is with the Photonics Research Group, Department of Electrical and Information Engineering, Politecnico di Bari, Via Orabona 4, 70126 Bari, Italy*
[2] *Richard Soref is with the University of Massachusetts at Boston, Boston, MA 02125 USA*
**francesco.deleonardis@poliba.it*

**Abstract:** We propose new Si-based waveguided Superlattice on Insulator (SLOI) platforms for high-performance electro-optical (EO) 2 x 2 and N x M switching and 1 x 1 modulation; including broad spectrum and resonant. We present a theoretical investigation, based on the tight-binding Hamiltonian, of the Pockels EO effect in the lattice-matched undoped $(GaP)_N/(Si_2)_M$, $(AlP)_N/(Si_2)_M$, $(ZnS)_N/(Si_2)_M$, $(AlN)_N/(3C-SiC)_M$, $(GaAs)_N/(Ge_2)_M$, $(ZnSe)_N/(GaAs)_M$, and $(ZnSe)_N/(Ge_2)_M$ wafer-scale short-period superlattices that are etched into waveguided networks of small-footprint Mach-Zehnder interferometers and micro-ring resonators to yield opto-electronic chips. The spectra of the Pockels $r_{33}$ coefficient have been simulated as a function of the number of the atomic monolayers for "non-relaxed" heterointerfaces. The large obtained $r_{33}$ values enable the SLOI circuit platforms to offer a very favorable combination of monolithic construction, cost-effective manufacturability, high modulation/switching speed, high information bandwidth, tiny footprint, low energy per bit, low switching voltage, near-IR-and-telecom wavelength coverage, and push-pull operation. By optimizing waveguide and clad and electrode dimensions, we obtained very desirable values of the $V_\pi L$ performance metric, in the range of 0.062 to 0.275 V×cm, portending a bright future for a variety of applications.

## 1. Introduction

This paper presents detailed analysis of waveguided electro-optical (EO) modulators and switches within seven new superlattice-on-insulator (SLOI) structures. The large-area SLOI wafer is a recently proposed [1],[2] CMOS-compatible optoelectronics (OE) platform that is projected to enable many key applications in computing, communications, sensing and quantum photonics. The initially uniform short-period superlattice stack is etched into a wafer-scale array of photonic integrated circuits (PICs) containing efficient Pockels networks. Ideally, the control electronics are integrated onto the same wafer, and thus this OE platform becomes the source of versatile, high-performance chips.

The SLOI wafer is believed to be compatible with high-volume foundry manufacture. We have proposed [2] to create the large-diameter foundry wafer by direct wafer bonding of a superlattice donor wafer to an oxidized silicon receiver wafer having for example, 300-mm diameter. The donor is cut back after bonding, leaving only the uniform SL stack. This $((A)_N/(B)_M)$ stack consists of *N* monolayers (MLs) of material *A* (group III-V) and *M* monolayers of material *B* (group IV or III-V).

This paper focusses on the engineering and optimization of the linear EO effect (the Pockels effect) in seven undoped, lattice-matched short-period superlattices (SPSLs) that provide fast, low energy, low-loss, waveguided EO modulation (broad spectrum or resonant modulators) as well as broad spectrum switching "meshes" (such as neuromorphic meshes, beam- steering meshes and N x M spatial routing matrices) operating at visible-light wavelengths in addition to telecom wavelengths. The results found here indicate that the SLOI chips will be fully competitive with Pockels chip formed from Lithium Niobate on Insulator and Barium Titanate on Insulator.

Our recent investigations [3] showed that very strong second-order and third-order nonlinear optical responses can be engineered in the GaP/Si SLOI, and second-order nonlinearity is the basis of the very large Pockels effects that we predict. By choosing the atomic layering of the SPSL properly, one wafer can feature excellent nonlinear and linear functions; that is, the SLOI PIC can give quantum light sources (or classical harmonics) as well as leading-edge Pockels devices.

The paper is organized as follows. Motivations for realizing Photonic integrated Circuits (PICs) based on SLOI are detailed in Section II. The theory background is reported in Section III. Numerical results about the $r_{33}$ calculations are detailed in Section IV.A. The large coefficients that are found are deployed in Section IV.B to demonstrate the potential of the proposed platform for inducing state-of-the-art EO modulators (both resonant and broad modulators), while Section IV.C quantifies the uniquely valuable SLOI potential for 2 x 2 crossbar switching. Combined EO nonlinear and PIC operation is discussed in Section V, while Section VI specifies a new method for integrating photodetectors on the SL waveguide. Finally, Section V summarizes the work.

## 2. Superlattice-on-Insulator Photonics

We focus on the following undoped SPSLs: $(GaP)_N/(Si_2)_M$, $(AlP)_N/(Si_2)_M$, $(ZnS)_N/(Si_2)_M$, $(AlN)_N/(3C-SiC)_M$, $(GaAs)_N/(Ge_2)_M$, $(ZnSe)_N/(GaAs)_M$, and $(ZnSe)_N/(Ge_2)_M$, in which the $N$ monolayers of the group III-V material and the $M$ monolayers of the group IV or III-V material are repeated periodically along the [111] growth direction ($z$ axis). Note that the single ML is comprised of a two-atom-thick layer (i.e cation Ga and anion P). Moreover, all atoms are on the sites of a zinc-blende lattice with lattice constant $a_L$ (under the perfect-matched condition). The $x$ and the $y$ coordinate axes are chosen along $[\bar{2}\ 1\ 1]$ and $[0\ \bar{1}\ 1]$. In order to realize waveguiding structures such as straight waveguides, Mach-Zehnder interferometers (MZIs) and micro-ring resonators (MRRs where optical mode confinement within the SPSL is needed. Thus, we propose to first grow the SPSL stack by molecular-beam epitaxy (MBE) [4] upon a large-area lattice-matched donor wafer such as 300-mm-diameter silicon or Ge-buffered silicon. The number of periods of layering, depending on the particular values of $a_L$, $N$ and $M$, is chosen in order to give a uniform stack thickness across the wafer, a thickness able to support fundamental optical modes at the operation wavelength (in the visible or near IR or mid IR). After that, the donor is bonded to an oxidized silicon receiver wafer having about 2-μm of top $SiO_2$. Next, the material above the SL is cut away to leave only the stack, thus creating a superlattice-on-insulator (SLOI) structure. Finally, an etching process is applied in order to create the fundamental photonic devices based on low-loss SPSL strip and rib waveguides, and the connected devices are configured into PICs with complex functionalities.

We recently quantified that the system Si-GaP can enable the possibility of realizing giant $\chi^{(2)}_{zzz}$ susceptibility. We proposed in [5] a multi-period stack of two doped asymmetric coupled quantum wells (ACQWs), where giant values of $\chi^{(2)}_{zzz}$ are induced by the combination of the dopant surface density and by the double resonance condition obtained via engineering the first three quantum-confined states. There, we explored the feasibility of realizing giant $\chi^{(2)}_{zzz}$ susceptibility in n-doped [3] and undoped [2] $(GaP)_N/(Si_2)_M$ SPSLs, where the coefficient $\chi^{(2)}_{ijk}(2\omega, \omega, \omega)$ was calculated considering the electron transitions inside the conduction band (CB) and between valence band (VB) and CB, respectively.

Because of the large susceptibilities $\chi^{(2)}$ and $\chi^{(3)}$ that were found in our initial SPSL, we expect to find large Pockels effects generally in SPSL structures, making the waveguided SLOI platform a good candidate for both classical and quantum applications. Indeed, efficient nonlinear effects such as Second Harmonic Generation, Spontaneous Parametric Down Conversion and Spontaneous Four Wave Mixing and efficient electro-optic modulation can be realized on the same technological platform. In the paragraphs below, we demonstrate that the Pockels coefficients of the SPSLs proposed here are lower than those of BaTiO$_3$, but much larger that the values recorded for LiNbO$_3$ and non-centrosymmetric semiconductors.

Moreover, within the monolithic semiconductor waveguides, we think the metrics of our EO-modulators can be very competitive with the metrics obtained in silicon PN carrier-depletion modulators.

For both classical and quantum applications, future advances in PICs will likely require a heterogeneous approach [6] that combines multiple materials in order to optimize light generation, optical amplification, modulation, switching, routing, qubit input/output (I/O) interfacing, passive splitting/combining, filtering, frequency translation, and photo-detection. In principle, our SLOI platforms could contribute to all of these areas (except perhaps lasing and gain).

## 3. Numerical Results

In this section, we present the numerical results on the Pockels coefficient obtained in superlattice material systems. Moreover, the performance metrics of both MZI and MRR electro-optic modulators based on SLOI waveguides will be derived as a function of the calculated $r_{33}$ coefficient. In this context, the cross-section of one waveguide and the geometrical parameters are sketched in Fig. 1 (a), where the SPSL $(A)_N/(B)_M$ represents any SL platform among $(GaP)_N/(Si_2)_M$, $(AlP)_N/(Si_2)_M$, $(ZnS)_N/(Si_2)_M$, $(AlN)_N/(3C-SiC)_M$, $(GaAs)_N/(Ge_2)_M$, $(ZnSe)_N/(GaAs)_M$, and $(ZnSe)_N/(Ge_2)_M$. In addition, Fig. 1 (b) shows a broad-spectrum 1 x 1 EO modulator based on a push-pull MZI architecture. A localized rib-waveguide structure in the push-pull region is assumed. The other waveguides are strips. Because of the planarized SiO$_2$ upper-cladding that covers all waveguides, the electrode configuration is assumed coplanar, where the signal electrode is placed on the top of the rib waveguide and the two ground electrodes are on the sides of each waveguide.

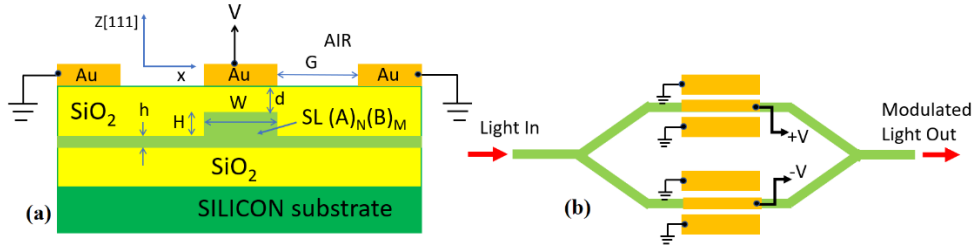

**Figure 1** (a) Cross section view of a Pockels-effect EO -waveguide modulator in the SLOI platform. (b) Top view of 1 x 1 EO amplitude-modulator based on the push-pull MZI architecture and the Fig. 1(a) waveguides.

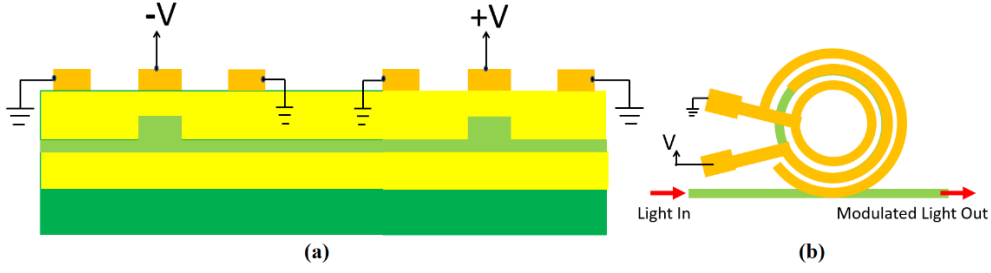

**Figure 2.** (a) Cross-section view of the active push-pull region in Fig. 1 (b) and in the 2 x 2 crossbar switch; (b) Top-view of the resonant bus-coupled waveguided SLOI MRR EO amplitude-modulator showing the co-planar ring electrodes employed upon the ring waveguide and the ring rib regions.

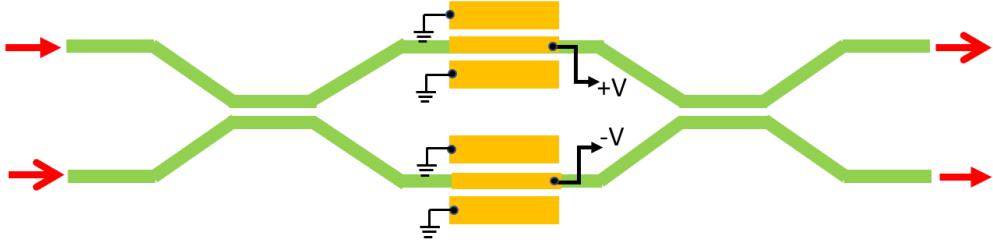

**Figure 3.** Top view of broad-spectrum 2 x 2 MZI crossbar switch in the SLOI platform. Electrodes on the localized push-pull region of Fig. 2 (a) are shown.

## *3.1 Pockels coefficient in Superlattice platform*

We perform the $r_{ijk} = r_{ijk}^{(el)} + r_{ijk}^{(ion)}$ calculations (see Methods section) for all combinations of $N$ and $M$ satisfying the condition $N+M = 3$ $N+M = 6$ and $N+M = 9$. The $C_{3v}$ point group is expected for a SPSL grown in the [1 1 1] direction [3].

The dominant electro-optic tensor element is $r_{33}$ (Voigt notation) and for this reason it will be object of the following analysis. Hereafter we will consider the $(GaP)_N/(Si_2)_M$ system as our reference SPSL platform, since very large second and third nonlinear effects (i.e $\chi_{zzz}^{(2)}(2\omega,\omega,\omega)$ and $\chi_{xxxx}^{(3)}(3\omega,\omega,\omega,\omega)$) have been demonstrated in our recent works [2],[3]. In this context, Fig. 4 shows the $r_{33}$ coefficient as a wavelength function for several values of $N$ and $M$. In each case the simulations have been performed for wavelength larger the inter-band edge wavelength (such as 822 nm), where the absorption losses are negligible.

We record a large value of the $r_{33}$ coefficient of 89.27 pm/V at 822.5 nm for the SPSL $(GaP)_2/(Si_2)_1$.

We have performed similar calculations for all of the above-mentioned superlattice platforms. The results of our investigations are summarized in Fig. 5, where for each SL we have reported only the cases giving the larger $r_{33}$ at the band-edge wavelength.

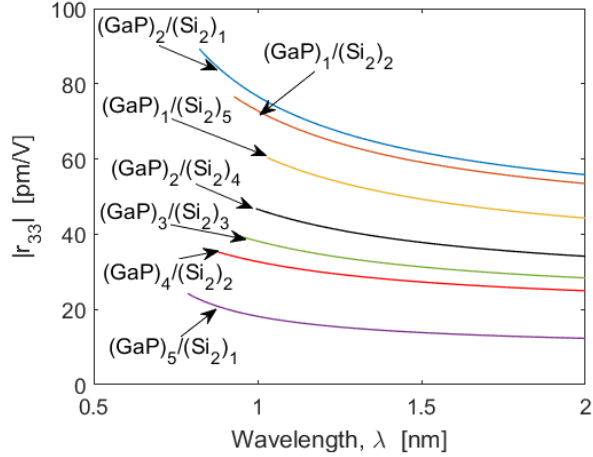

**Figure 4.** Pockels $|r_{33}|$ coefficient as a function of the wavelength, $\lambda$, for different $(GaP)_N/(Si_2)_M$ SPSLs.

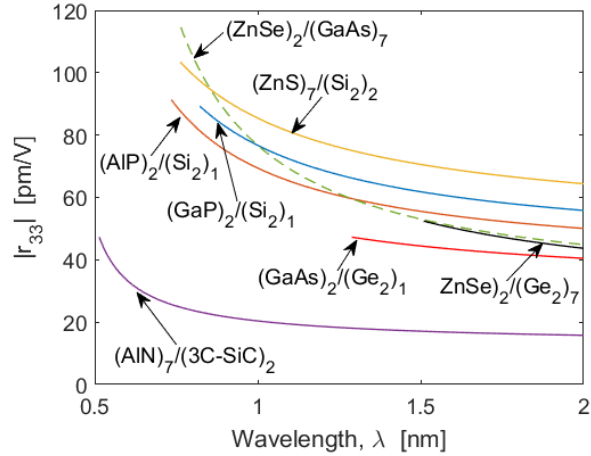

**Figure 5.** Pockels $|r_{33}|$ coefficient as a function of the wavelength, $\lambda$, for different optimized SL platforms.

It is interesting to note that very large values of $r_{33}$=103.5 pm/V and 114.7 pm/V have been obtained for $(ZnS)_7/(Si_2)_2$ and $(ZnSe)_2/(GaAs)_7$, respectively. Moreover, all the curves indicate that values of $r_{33}$ larger than those of the bulk semiconductors can obtained, enabling the SLOI platform to be a good candidate for realizing high performance EO modulators for wavelengths ranging from the 500-nm visible to the near-infrared and through the mid-infrared region.

Figure 6 shows the larger values of the $r_{33}$ coefficient, calculated at 1550 nm and for all the SL platforms considered.

It is worth outlining that in the all $r_{33}$ calculations, both the Hamiltonian $H(\mathbf{k})$ and the dipole matrix elements are calculated taking into account the valence band offset parameter, $\Delta E_v$ (see Methods section). The $\Delta E_v$ values used in the simulations are listed in Table I for all the considered SL platforms.

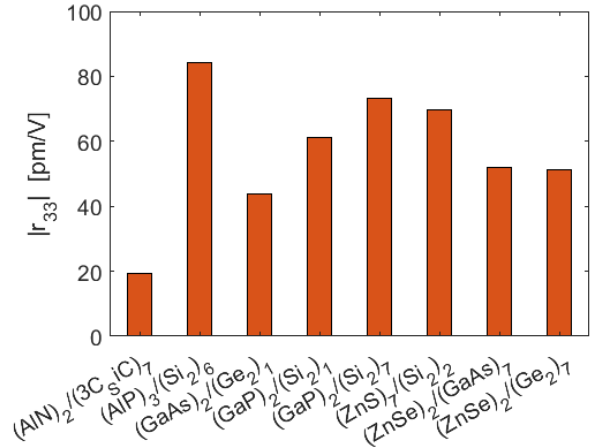

**Figure 6**. Pockels $|r_{33}|$ coefficient calculated at telecom $\lambda$=1550 nm, for different optimized SL platforms.

### *3.2 EO 1 x 1 Modulators based on the SLOI Platform.*

Due to the electro-optical modification of the refractive index in a Pockels material, light undergoes a controllable modulation in its optical phase as it travels through the medium. This is electro-refraction (phase shifting) without electro-absorption. This shifting is utilized in electro-optic devices, where the device architecture is set on the basis of the specific application. Commonly, broad-spectrum Mach-Zehnder interferometers in the 1 x 1 configuration are used in order to induce amplitude modulation. In particular, MZIs allow push-pull operation, where the voltage is applied across both arms with an equal magnitude but with opposite sign with respect to each other. Then, the interferometer arms experience opposite phase shift, inducing a $\pi$-phase difference over a modulation length ($L$) scaled by a factor of 2 when compared with the single-arm operation at the same driving voltage. Generally, the overall performance of EO devices not only depends on the Pockels coefficients of the but also on the confinement of light to the active electro-optic material and the electro-optic field overlap. In this sense, Fig. 7 shows both the spatial distribution of $|\boldsymbol{E}|^2$ for the fundamental TM mode and the applied static electric field $E_{RF}$ (black arrows). Our simulations are based on full-vectorial FEM calculations where a multiphysics approach has been adopted. The FEM electromagnetic module that is used to evaluate the optical mode distributions inside the active region of the EO modulator, works together with the FEM static module in order to simulate the distribution of the applied static electric field and the capacitance per unit length ($C/L$) of the EO structure. Under this scenario, the overlap between the optical and static electric fields is calculated in order to better evaluate the product $V_\pi L$ (see Eqs. (1)-(2)).

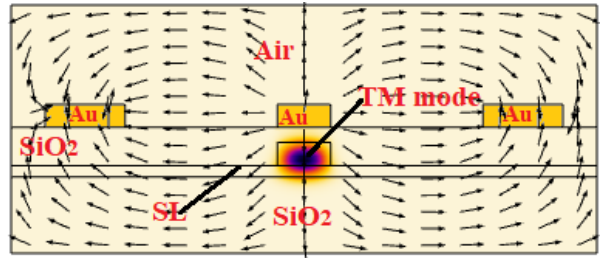

**Figure 7.** Simulated spatial distribution of the fundamental TM mode (color map) and applied static electric field (black arrows). In the simulation: $W$=700 nm, $H$=300 nm, $h$=150 nm, $d$=500 nm and SPSL: $(GaP)_2/(Si_2)_1$

Ideally, in Fig. 7, the RF field should be vertical in the core of the SL waveguide to maximize the $r_{33}$ response, but the field direction in practice deviates from the ideal orientation.

The half-wave voltage-length product $V_\pi L$ has emerged as a useful figure of merit (FOM). An ultimate goal of a modulator design is to achieve the smallest possible $V_\pi L$, resulting in the smallest footprint with the least required voltage. Moreover. a reduced $V_\pi L$ product leads to further performance benefits with respect to modulation bandwidth and energy consumption. A shorter interaction length induces a smaller device capacitance $C$ (due to a decreased electrode area), which in turn determines larger bandwidth $f_{3dB}$ and lower electric power consumption $E_{bit}$ according to: $f_{3dB} = 1/2\pi RC$ and $E_{bit} = CV^2/4$, respectively.

Assuming a symmetrical push-pull MZI (see Fig. 1 (b)), the $V_\pi L$ is given by [15]:

$$V_\pi L = \frac{\lambda n_{eff}}{2\pi n_z^4 \Gamma} \quad (1)$$

where $n_{eff}$ is the effective index of the waveguide mode (TM in our case) and $n_z$ the material index of SL structure (see Fig. 1 (a)). The term $\Gamma$ is an overlap integral parameter given by [15]:

$$\Gamma = \frac{\iint [E_{RF}(x,y)/V] r_{33}(x,y) |E(x,y)|^2 \, dxdy}{\iint |E(x,y)|^2 \, dxdy} \quad (2)$$

where $E_{RF}/V$ is the RF frequency electric field per volt applied to the electrodes, and $E$ is the electric field of the fundamental TM optical mode.

A further figure of merit is the optical insertion loss IL, induced by the loss coefficient $\alpha$ inside the integrated waveguide. Our expectation is that the IL of any SLOI passive (non electroded) waveguide, for all seven cases, will be well below 1 dB/cm, although unwanted surface-scattering loss could affect "overall IL" if the fabrication process is not well controlled.

For EO modulators and switches we need to minimize the electrode-induced losses ($\alpha_{el}$), in order to keep the total propagation loss below 1 dB/cm. Our preliminary simulations indicate that $\alpha_{el}$ is weakly dependent on the electrode gap ($G$), but it is strongly influenced by the geometrical parameters $d$ (see Fig. 1 (a)). In this context, Fig. 8 plots the contribution $\alpha_{el}$ as a function of the parameter $d$, for the SPSL $(GaP)_2/(Si_2)_1$ and assuming: $W$=700 nm, $H$=300 nm (corresponding to 320 SL periods), $h$=150 nm (corresponding to 159 SL periods), $G$=2000 nm, and $\lambda$=822.5 nm.

The plot clearly indicates that values of $d$ larger than 500 nm are required in order to keep $\alpha_{el}$ lower than 0.1 dB/cm (considered negligible with respect to the contributions induced by both the fabrication process and surface roughness).

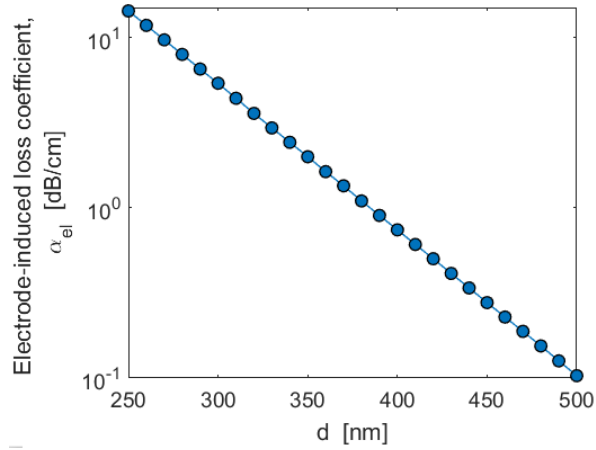

**Figure 8.** Simulated electrode-induced loss coefficient, $\alpha_{el}$, as a function of the parameter $d$. In the simulation: $W$=700 nm, $H$=300 nm, $h$=150 nm, $G$=2000 nm, $\lambda$=822.5 nm, and SPSL: $(GaP)_2/(Si_2)_1$

Under this scenario, the figures of merit of the MZI push-pull modulator, sketched in Fig. 1 (b), are shown in Fig. 9, where the product $V_\pi L$ and $C/L$ are plotted as a function of the electrode gap, $G$.

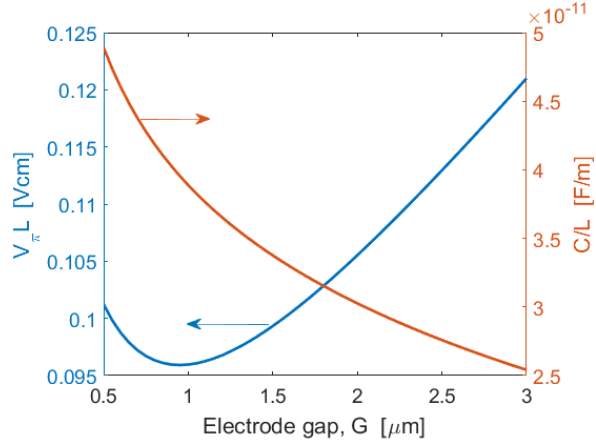

**Figure 9.** Simulated $V_\pi L$ and $C/L$ as a function of the electrode gap $G$. In the simulation: $W$=700 nm, $H$=300 nm, $h$=150 nm, $d$=500 nm, $\lambda$=822.5 nm, and SPSL: $(GaP)_2/(Si_2)_1$

The plot records a minimum of $V_\pi L$=0.096 Vcm at $G$=1000 nm. Moreover, the parameter $C/L$ decreases, increasing the electrode Gap. For an example setting $G$=2000 nm, as a trade-off choice, we record the following performances for a modulation length $L$=2 mm [16]: $V_\pi$=0.53 V, $C$=60.52 fF, $f_{3dB}$=52.6 GHz, and $E_{bit}$=4.21 fJ/bit. With short modulation length, the electrodes can be treated as lumped elements driven as capacitors. The modulation bandwidth is, thus, limited by the product of the capacitance ($C$) and total resistance ($R$), currently limited by the impedance of the network analyzer drive (50 Ω). However, we can obtain modulation bandwidth much higher value (>100 GHz) using a travelling wave design, where the bandwidth is influenced by the velocity mismatch between the optical and RF signals and by the metal RF loss [15]. However this approach is not considered here, since we are interested in a modulator design minimizing the footprint.

Using the Fig. 9 procedure, and taking L = 1 mm, we performed a two-parameter minimization of the $V_\pi L$ for all seven platforms, determining for each the combination of $d$ and $G$ values that gave the optimum FOM, with the $d$ value being chosen under the assumption of 1 dB/cm loss for the active region of the modulator or switch, The results are displayed in Table I. The 0.062 to 0.275 FOMs are competitive with those of LiNbO$_3$ and BaTiO$_3$. (Table II of [17]). The switching voltage here $V_s$ is the same as the $V_\pi$ voltage. The $V_s$ is found to range from 0.68 V to 2.75 V at the wavelength shown, and this will yield a high extinction ratio, such as 15 dB in the modulators. If we could accept 7 dB of extinction, for example, then the modulation voltage would be about one-half of what is listed. Note that the IL is quite small in each case. Bandwidths exceed 72 GHz and energy per bit is in the fJ/bit range.

**TABLE I**

Figures of Merit of EO MZI push-pull 1 x 1 Modulators and 2 x 2 switches based on different SLOI Platforms. The active length is 1 mm for modulators and switches. The assumed insertion loss of the modulators and the switches is 1 dB per cm of active length. Figure 1 (a) dimensions $d$ and $G$ are given for optimized $V_\pi L$.

| SPSLs | W×H [nm]×[nm] | λ [nm] | $\Delta E_v$ [eV] | $|r_{33}|$ [pm/V] | $V_\pi L$ [V×cm] | $V_s$ [V] | C/L [F/m] | $E_{bit}$ [fJ/bit] | $f_{3dB}$ [GHz] |
|---|---|---|---|---|---|---|---|---|---|
| (GaP)$_2$/(Si$_2$)$_1$  $d$ =385 nm; $G$ =0.85 μm | 700×300 | 822.5 | 0.24 [3] | 89.27 | 0.08 | 0.8 | 4.105×10$^{-11}$ | 6.57 | 77.54 |
| (AlP)$_2$/(Si$_2$)$_1$  $d$ =355 nm; $G$ =0.80 μm | 700×300 | 735.1 | 0.24 [3] | 91.33 | 0.068 | 0.68 | 4.193×10$^{-11}$ | 4.86 | 75.91 |
| (ZnS)$_7$/(Si$_2$)$_2$  $d$ =410 nm; $G$ =1.0 μm | 700×400 | 762.2 | 0.70 [18] | 103.5 | 0.10 | 1,00 | 3.92×10$^{-11}$ | 9.9 | 81.14 |
| (ZnSe)$_2$/(GaAs)$_7$  $d$ =360 nm; $G$ =0.80 μm | 700×300 | 763.8 | 0.72 [19] | 114.7 | 0.062 | 0.62 | 4.214×10$^{-11}$ | 4.08 | 75.53 |
| (ZnSe)$_2$/(Ge$_2$)$_7$  $d$ =392 nm; $G$ =1.0 μm | 900×600 | 1510 | 1.12 [20] | 52.4 | 0.275 | 2.75 | 4.40×10$^{-11}$ | 83.38 | 72.35 |
| (GaAs)$_2$/(Ge$_2$)$_1$  $d$ =395 nm; $G$ =0.95 μm | 900×550 | 1290 | 0.31 [21] | 47.24 | 0.244 | 2.44 | 4.38×10$^{-11}$ | 65.31 | 72.68 |
| (AlN)$_7$/(3C − SiC)$_2$  $d$ =391 nm; $G$ =1.1 μm | 400×220 | 513.8 | 0.65 [22] | 47.29 | 0.12 | 1.2 | 3.035×10$^{-11}$ | 11.03 | 104.9 |

At this step we will analyze the performances of the EO modulators based on MRR architecture (see Fig. 2 (b)) taking again the 1 dB/cm assumption. In this context, we need to apply an external voltage ($V$) that is large enough to cause a resonance shift of $\Delta \lambda = 1 \times \delta \lambda$ or $\Delta \lambda = 2 \times \delta \lambda$, where $\delta \lambda$ represents the 3-dB resonance linewidth. Moreover, the total switching delay is defined as $\tau_{switch} = \tau_{RC} + \tau_{photon}$ where $\tau_{RC}$ is the $RC$ constant of the electrodes, and $\tau_{photon} = \lambda_0 Q/2\pi c_0$ is the photon lifetime in the micro-ring resonator. Here, $c_0$, $\lambda_0$ and $Q$ are the vacuum light velocity, the ring resonance wavelength and the cavity quality factor, respectively.

According to static FEM simulations (see $C/L$ in Table I), we expect that for short cavity $\tau_{photon}$ is generally much longer than the $RC$ constant. As a result, the bandwidth is limited by the photon lifetime. In Fig. 10 we show the applied voltage needed to satisfy the condition $\Delta\lambda = 1 \times \delta\lambda$ and the photon lifetime-induced bandwidth as a function of the cavity quality factor, for different SLOI platforms. In the simulations we have assumed a MRR having a ring radius of 50 µm [16] and the geometrical cross-section parameters as listed in Table I. Although we have performed the simulations for all the considered platforms, not all curves are plotted in Fig. 10 due to the overlapping among them. Here, the calculated bandwidth induced by the $RC$ constant ranges from a minimum of 230 GHz for $(ZnSe)_2/(Ge_2)_7$ to a maximum of 334 GHz for $(AlN)_2/(3C-SiC)_7$, confirming that the bandwidth of electro-optic MRR modulator is limited by the photon lifetime. If fabrication of the MRR gives smooth waveguide walls, a $Q$ of 20,000 should be achievable and in that case Fig. 10 indicates modulation voltage in the 1.17 to 2.77 V range. If higher-Q is attained, the required voltage is 0.5 to 1.0V.

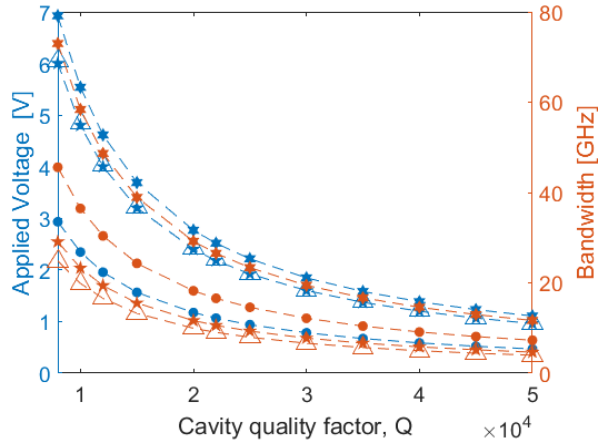

**Figure 10**. Applied MRR voltage, satisfying the conditions: $\Delta\lambda = 1 \times \delta\lambda$ and the photon-lifetime-induced bandwidth as a function of the cavity quality factor. Hexagram: $(AlN)_2/(3C-SiC)_7$; Pentagram: $(GaAs)_2/(Ge_2)_1$; Triangle: $(ZnSe)_2/(Ge_2)_7$; Circle: $(GaP)_2/(Si_2)_1$.

## *3.3 Title Broad Spectrum EO 2 x 2 MZI Crossbar Switches in the SLOI Platform.*

To move toward large-scale integration of 2 x 2 switches in N x N switching matrices we need to reduce the footprint of the device by selecting an active length of 50 µm or 200 µm. This would increase the switching voltage requirement on $V_s$ in a tradeoff condition. One way to constrain or reduce $V_s$ is to decrease the thickness of the upper oxide, that is, reducing $d$, which in turn increases IL via electrode loss.

However, the smaller active length means that a higher loss per unit length is acceptable. Hence we now chose 10 dB/cm for the switch investigation (see Fig. 3).

Taking initially L = 200 µm, we then performed a set of simulations where the $V_\pi L$ minimization was obtained by proper choice of $d$ and $G$. Our results are presented in Table II for selected best-performing platforms at the wavelength-of-operation shown.

The simulation work in Table II is summarized as $V_\pi L$ in the range 0.051 to 0.204 Vxcm, cross-bar voltage 2.57 to 10.22 Volts, switching time less than 1 ps, switching energy 15 to 280 fJ/bit, and capacitance around 10 fF. At the telecom wavelength, the AlP/Si platform gives full switching at 9.8 Volts. Regarding large scale integration for matrix switching, we note that the active length L could be reduced to 50 μm for the ZnSe/GaAs platform at the 764 nm wavelength, and in that case the cross-bar applied voltage would be 10.4 Volts.

The Tables I and II modulation and switching results given here indicate that our EO SLOI platforms are quite competitive with the leading state-of-the-art Pockels EO waveguided circuit platforms-- ferroelectric and poled organic polymer-- in terms of Pockels coefficient value, switching voltage, footprint, speed, bandwidth, energy required and $V_\pi L$.

**TABLE II**
Figures of Merit of EO MZI broadband push-pull 2 x 2 Crossbar Switch based on different SLOI Platforms
The active length is 200 um for these Fig. 3 switches. The assumed insertion loss of the switches
is 10 dB per cm of active length. Figure 1 (a) dimensions $d$ and $G$ are given for optimized $V_\pi L$.

| SPSLs | $W \times H$ [nm]×[nm] | $\lambda$ [nm] | $V_\pi L$ [V×cm] | $V_s$ [V] | $C$ [fF] | $E_{bit}$ [fJ/bit] | Switch time [ps] |
|---|---|---|---|---|---|---|---|
| $(GaP)_2/(Si_2)_1$ $d$ =268 nm; $G$ =0.65 μm | 700×300 | 822.5 | 0.065 | 3.27 | 8.98 | 23.98 | 0.449 |
| $(GaP)_2/(Si_2)_7$ $d$ =295 nm; $G$ =0.7 μm | 1200×700 | 1550 | 0.204 | 10.22 | 10.75 | 280.7 | 0.537 |
| $(AlP)_2/(Si_2)_1$ $d$ =251 nm; $G$ =0.7 μm | 700×300 | 735.1 | 0.057 | 2.83 | 8.78 | 17.64 | 0.439 |
| $(AlP)_3/(Si_2)_6$ $d$ =315 nm; $G$ =0.75 μm | 1200×700 | 1550 | 0.196 | 9.8 | 10.52 | 252.9 | 0.525 |
| $(ZnS)_7/(Si_2)_2$ $d$ =287 nm; $G$ =0.80 μm | 700×400 | 762.2 | 0.082 | 4.12 | 8.47 | 36.05 | 0.423 |
| $(ZnSe)_2/(GaAs)_7$ $d$ =253 nm; $G$ =0.65 μm | 700×300 | 763.8 | 0.051 | 2.57 | 9.05 | 15.0 | 0.453 |

## 4. Multi-Function Superlattice Circuits

We have detailed the many Pockels possibilities for the SLOI PICs, and we wish to point out that a given Pockels PIC—any of the above seven SLOI PICs—also offers strong second-order and third-order nonlinear optical (NLO) response. In other words, a given PIC will perform NLO and EO functions at the same time, along with giving a complicated passive-waveguide circuit. Detector arrays are easily formed there as follows.

## 5. Integrated Photodetectors that use One Superlattice Semiconductor

The SPSL is a "designed semiconductor" –a new, man-made semiconductor comprised of well-known semiconductors A and B, and the SPSL bandgap Eg(S) is the approximate average of Eg(A) and Eg(B). If we assume that Eg(B) is the smaller gap, then semiconductor B by itself would absorb all of the on-chip photons whose hv energy is Eg(S) < hv < Eg(B). We can put that absorption to use by growing a lattice-matched layer of B upon a strip waveguide in the A/B SPSL circuit, and by adding lateral P and N doping of the B rectangular upper-clad region. In that case, we have thereby formed a lateral PIN photodiode with evanescent-wave coupling to light traveling within the strip, since the refractive index of B is higher than that of the SPSL, causing automatic upward leakage of the guided mode into the PIN-B upper cladding volume. In other words, it will be easy to fabricate monolithic detector arrays in the SLOI PIC, and the detector material will be silicon for ZnS/Si, AlP/Si and GaP/Si platforms. It will be Ge for ZnSe/Ge and GaAs/Ge platforms, and it will be GaAs for the ZnSe/GaAs platform. Employing a similar procedure, monolithic avalanche photodiodes can be constructed on the SL waveguides in a lateral APD geometry.

## 6. Conclusions

In this paper, a physics-and-engineering procedure has been performed in order to investigate the Pockels coefficient in $(GaP)_N/(Si_2)_M$, $(AlP)_N/(Si_2)_M$, $(ZnS)_N/(Si_2)_M$, $(AlN)_N/(3C-SiC)_M$, $(GaAs)_N/(Ge_2)_M$, $(ZnSe)_N/(GaAs)_M$, and $(ZnSe)_N/(Ge_2)_M$. short-period- and ultra-short-period superlattices. In this context, general physical aspects have been investigated by means of the empirical $sp^3s^*$ tight-binding method, by determining the features of the electronic-and-ionic structure and the influence of the monolayers number upon $\chi^{(2)}_{zzz}(\omega, \omega, 0)$. We have explored large $r_{33}$ coefficients in the proposed SL structures in the 500 nm to 2000 nm wavelength range, demonstrating that efficient and "competitive" electro-optic modulation can be generated in both integrated broad-spectrum Mach-Zehnder interferometers and micro-ring resonators. We investigated also broadband 2 x 2 MZI cross-bar spatial routing switches for N x N routing networks. The dominant Pockels coefficient in these SLOI platforms allows push-pull switching in all MZIs together with small footprint, low voltage, high speed, and energy efficiency.

Looking towards the future development of "on insulator" silicon-based photonic integrated circuits (which might include on-chip control electronics), the seven SLOI platforms proposed here feature a combination of desireable features: (1) a network of low-loss passive SL strip waveguides-and-components, (2) wide coverage of operation wavelengths, (3) strong nonlinear optical susceptibilities, (4) high performance Pockels-effect EO 1 x 1 modulators and 2 x 2 switches that offer a combination of excellent metrics, (5) simultaneous EO and NLO operation, and (6) monolithic lattice-matched on-chip SL waveguide-integrated photodetectors utilizing one of the two SL materials.

A wafer-bonding technique could be used to create the SLOI circuits just described. The theoretical work in this paper suggests that the undoped lattice-matched short-period superlattice platforms, in their seven-fold variety, could in the future offer cost-effective solutions to next-generation data center interconnections and to a variety of other important applications.

## 7. Methods

In this section we describe the mathematical modeling for the calculation of the Pockels coefficient, generated in short-period superlattices based on $(GaP)_N/(Si_2)_M$, $(AlP)_N/(Si_2)_M$, $(ZnS)_N/(Si_2)_M$, $(AlN)_N/(3C-SiC)_M$, $(GaAs)_N/(Ge_2)_M$, $(ZnSe)_N/(GaAs)_M$, and $(ZnSe)_N/(Ge_2)_M$, technological "on insulator" platforms. Microscopically, the Pockels coefficient is a tensor, $r_{ijk}$, that relates the change in the component $\varepsilon_{ij}$ of the inverse optical dielectric tensor of a non-centrosymmetric crystal to a static (or low frequency), external electric field applied in the k-direction ($E_k$), according to the following equation [7,8]:

$$\Delta\left(\varepsilon^{-1}\right)_{ij} = \sum_k r_{ijk} E_k \tag{3}$$

by neglecting any modification of the unit cell shape due to the inverse piezoelectric effect, the EO tensor is only the sum of an electronic ($r_{ijk}^{(el)}$) contribution and an ionic ($r_{ijk}^{(ion)}$) contribution. The electronic contribution is due to an interaction of the electric field with the valence electrons when considering the ions artificially clamped at their equilibrium positions. It is proportional to the second-order nonlinear optical susceptibility $\chi_{ijk}^{(2)}(\omega,\omega,0)$, according to the Eq. (4) [8-10]:

$$r_{ijk}^{(el)} = -\frac{8\chi_{ijk}^{(2)}(\omega,\omega,0)}{n_i^2 n_j^2} \qquad (4)$$

where $n_i$ and $n_j$ are the principal refractive indices of the SPSL.

The calculations of $n_i$, $n_j$ and $\chi_{ijk}^{(2)}(\omega,\omega,0)$ are performed within the tight-binding (TB) framework. It describes the SPSL system by means of a real-space Hamiltonian-matrix function ($H(\mathbf{k})$), of a reduced number of parameters, resulting in a reduction of the computational costs from the ab-initio methods. The motivations for this choice and the calculation details are well presented in our previous works [3], where the $H(\mathbf{k})$ Hamiltonian of two different alternating zincblende crystals, labelled $ca$ and $CA$ (where c (C) and a (A) are the cation and the anion, respectively) is represented as a block matrix. The number of block matrices is strictly related to the MLs numbers $N$ and $M$. The block matrices are 10x10 matrices denoted as $H_{ca(CA)}$, $G_{ca(CA)}$ and $F_{ca(CA)}$ and representing the intramaterial interaction for $(ca)_N$ and $(CA)_M$, respectively. Moreover, the matrices $G_i$ and $F_i$ ($i = 0, 1$) are included in order to describe the intermaterial interactions. In this context, we first calculate the electronic structure of the SPSL by diagonalizing the Hamiltonian $H(\mathbf{k})$ and consequently we calculate the dipole matrix elements needed to evaluate the terms $n_i$, $n_j$ and $\chi_{ijk}^{(2)}(\omega,\omega,0)$ [2],[7].

By definition, the dipole matrix element is given by: $\langle n,\mathbf{k}|\mathbf{u}\cdot\mathbf{r}|m,\mathbf{k}\rangle = \mathbf{u}\cdot\langle n,\mathbf{k}|\mathbf{r}|m,\mathbf{k}\rangle$, where $\mathbf{u}$ is the polarization vector of the electric field and $\mathbf{r}$ is the vector position operator. Here $n$ and $m$ denote the band index, running in either VB or CB, and $\mathbf{k}$ indicates the wavevector. As detailed in [11-14], in $k$-space the position operator is proportional to the gradient with respect to k of the Hamiltonian $H(\mathbf{k})$:

$$\mu_{nm} = \langle n,\mathbf{k}|\mathbf{r}|m,\mathbf{k}\rangle = \frac{j}{(E_m(\mathbf{k}) - E_n(\mathbf{k}))}\langle n,\mathbf{k}|\nabla_k H(\mathbf{k})|m,\mathbf{k}\rangle \qquad (5)$$

where $E_i(\mathbf{k})$ represents the energy at the $\mathbf{k}$ point for the i-th band.

The ionic contribution to the EO response depends upon the relaxation of the atomic positions due to the applied quasistatic electric field, which, in turn, induces the variations on the tensor elements $\varepsilon_{ij}$. Rigorously speaking, the term $r_{ijk}^{(ion)}$ is related to the Raman susceptibility tensor, as described in [9]. However, it requires infrared Raman spectroscopy in order to evaluate the phonon mode dispersion curves. Due to the novelty of the superlattice platforms adopted here, no experimental data are present in literature. Thus to meet the purpose of predicting the properties of new nanocomposite materials such as the SPSL proposed here, the theory is required to employ as few physical parameters as possible. In this context, we adopt the semiclassical approach based on the energy band diagram, the dielectric theory and the concepts of bond charge and effective ionic charge. As indicated in [8], the band gap $E_g$ (evaluated here by means of the TB method [3]) can be decomposed into the covalent ($E_h$) and ionic ($E_C$) parts. The contribution $E_h$ depends only on the internuclear spacing, while the ionic gap $E_C$ is given by a screened Coulomb interaction. On the basis of the estimations of both $E_h$ and $E_C$, we calculate the contribution $r_{ijk}^{(ion)}$ by means of Eq. (14.44) of Ref. [8], which accounts reasonably well for both the magnitude and sign of $r_{ijk}^{(ion)}$ in zinc blende and wurtzite crystals. We are aware that the above theory is based on a variety of simplifying assumptions. However, since the foundations of this theory are empirical, some of the physical parameters involved in the model can be better set on the basis of future experimental measurements.

**Acknowledgments.** R. Soref thanks the Air Force Office of Scientific Research under Grant FA9550-21-1-0347.